\documentstyle[11pt,newpasp,twoside,epsf]{article}
\def\edcomment#1{\iffalse\marginpar{\raggedright\sl#1\/}\else\relax\fi} 
\reversemarginpar 
 
\begin{document} 
\title{Polar Ring Galaxies and the Tully-Fisher relation: implications
 for the dark halo shape} 
\author{Magda Arnaboldi} 
\affil{INAF, Astronomical Observatory of Turin} 
\author{Enrica Iodice} 
\affil{INAF, Astronomical Observatory of Capodimonte} 
\author{Frederick Bournaud \& Francoise Combes} 
\affil{LERMA, Observatory of Paris} 
\author{Linda S. Sparke} 
\affil{Dept. of Astronomy, University of Wisconsin} 
\author{Wim van Driel} 
\affil{GEPI, Observatory of Paris} 
\author{Massimo Capaccioli} 
\affil{INAF, Observatory of Capodimonte} 
 
\begin{abstract} 
We have investigated the Tully-Fisher relation for Polar Ring Galaxies
(PRGs), based on near infrared, optical and \mbox{H\,{\sc i}} data
available for a sample of these peculiar objects.  The total K-band
luminosity, which mainly comes from the central host galaxy, and the
measured \mbox{H\,{\sc i}} linewidth at $20\%$ of the peak line flux
density, which traces the potential in the polar plane, place most
polar rings of the sample far from the Tully-Fisher relation defined
for spiral galaxies, with many PRGs showing larger \mbox{H\,{\sc i}}
line-widths than expected for the observed K band luminosity.  This
result is confirmed by a larger sample of objects, based on B-band
data.  This observational evidence may be related to the dark halo
shape and orientation in these systems, which we study by numerical
modeling of PRG formation and dynamics: the larger rotation velocities
observed in PRGs can be explained by a flattened polar halo, aligned
with the polar ring.
\end{abstract} 
 
\section{Introduction} 
 Polar Ring Galaxies (PRGs) are peculiar objects composed of a central
spheroidal component, the host galaxy, surrounded by an outer ring,
made up of gas, stars and dust, which orbits nearly perpendicular to
the plane of the gas-poor central galaxy (PRGC; Whitmore et al. 1990).
Previous papers (Arnaboldi et al. 1995, 1997; Iodice
et al.  2002a, 2002b, 2002c) found that even where the morphology of
the host galaxy resembles that of an early-type system, PRGs show many
similarities with late-type galaxies.  The PRGs are characterized by a
large amount of neutral hydrogen (\mbox{H\,{\sc i}}), always
associated with the polar structure (Schechter et al. 1984; van Gorkom
et al. 1987; Arnaboldi et al. 1997), and by a gas-to-total luminosity
ratio in the B-band typical of late-type galaxies.

By exploring the properties of the host galaxy and ring in the optical
and NIR, for a sample of PRGs, Iodice et al.  (2002a, 2002b, 2002c)
found that the connection with spirals is tighter. 
The Tully-Fisher relation (TF) is the most important scaling relation
for disks (Tully \& Fisher, 1977): this is an empirical relationship
between the disk rotational velocity ($V_{rot}$) and its absolute
luminosity ($L$), where $L\propto V_{rot}^4$, approximately.  In the
past few years, several studies have asserted the validity of the TF
relation for some classes of disk galaxies which show different
photometric and kinematical properties with respect to `classical',
high-surface-brightness spiral galaxies (Matthews, van Driel \&
Gallagher 1998a, 1998b; McGaugh et al. 2000; Chung et al. 2002).
These latest developments indicate that the TF relation is probing a
very close liaison between the dark halo parameters and the total
quantity of baryons in galaxies: the dark halo, which is responsible
for the \mbox{H\,{\sc i}} linewidth and the flat rotation curve in the
outer regions of a disk, is tuned to the total amount of baryons in
the luminous component.

In PRGs, the \mbox{H\,{\sc i}} linewidth ($\Delta V$) measures the
dynamics along the meridian plane, which is dominated by the dark
matter, while the baryons are nearly equally distributed between the
host galaxy and the polar ring.  We wish to investigate the position
of the PRGs in the $\log(\Delta V)-L$ plane, and study via N-body
simulations of 3-D systems whether the dark halo shape may influence
their position in the $\log(\Delta V)-L$ plane, with respect to the TF
relation of bright disks.  The question of the dark halo shape is
important {\it i}) to constrain dark matter models, through
cosmological simulations (Navarro, Frenk \& White 1996, 1997; Bullock
et al. 2001) which predict the distribution of the halo shapes and the
universal radial dependence of the dark matter distribution; {\it ii})
to give hints on the nature of dark matter (see Combes 2003 as a
review); and furthermore {\it iii)} the dark halo properties in PRGs
can give important constraints on the formation scenarios for these
peculiar objects, which is still an open issue (see Iodice et
al. 2002a, 2002c, and references therein).

\section{Observations}
 
New near-infrared J, H and Kn images are available for a sample of
PRGs (Iodice et al. 2002a, 2002b, 2002c), which are selected from the
PRGC, and for ESO~235-G58.  The B band magnitude is known for many
PRGs (Van Driel et al. 2002, 2003; Gallagher
et al. 2002), and the \mbox{H\,{\sc i}} integrated line profile data were
obtained from several published hydrogen observations of PRGs, carried
out by e.g. Richter et al. (1994), van Gorkom et al. (1987), van Driel
and collaborators (2000, 2002, 2003), with several radio telescopes.
 
To derive the TF relation for normal disk galaxies, we will use the very 
large and detailed dataset available in the I-band from Giovanelli et al. 
(1997). We estimate the B band magnitude for spiral disks in
the sample from Giovanelli et al. using the observational relation
between morphological type index and the $B-I$ colors (de Jong
1996).

\begin{figure}[h]
\plottwo{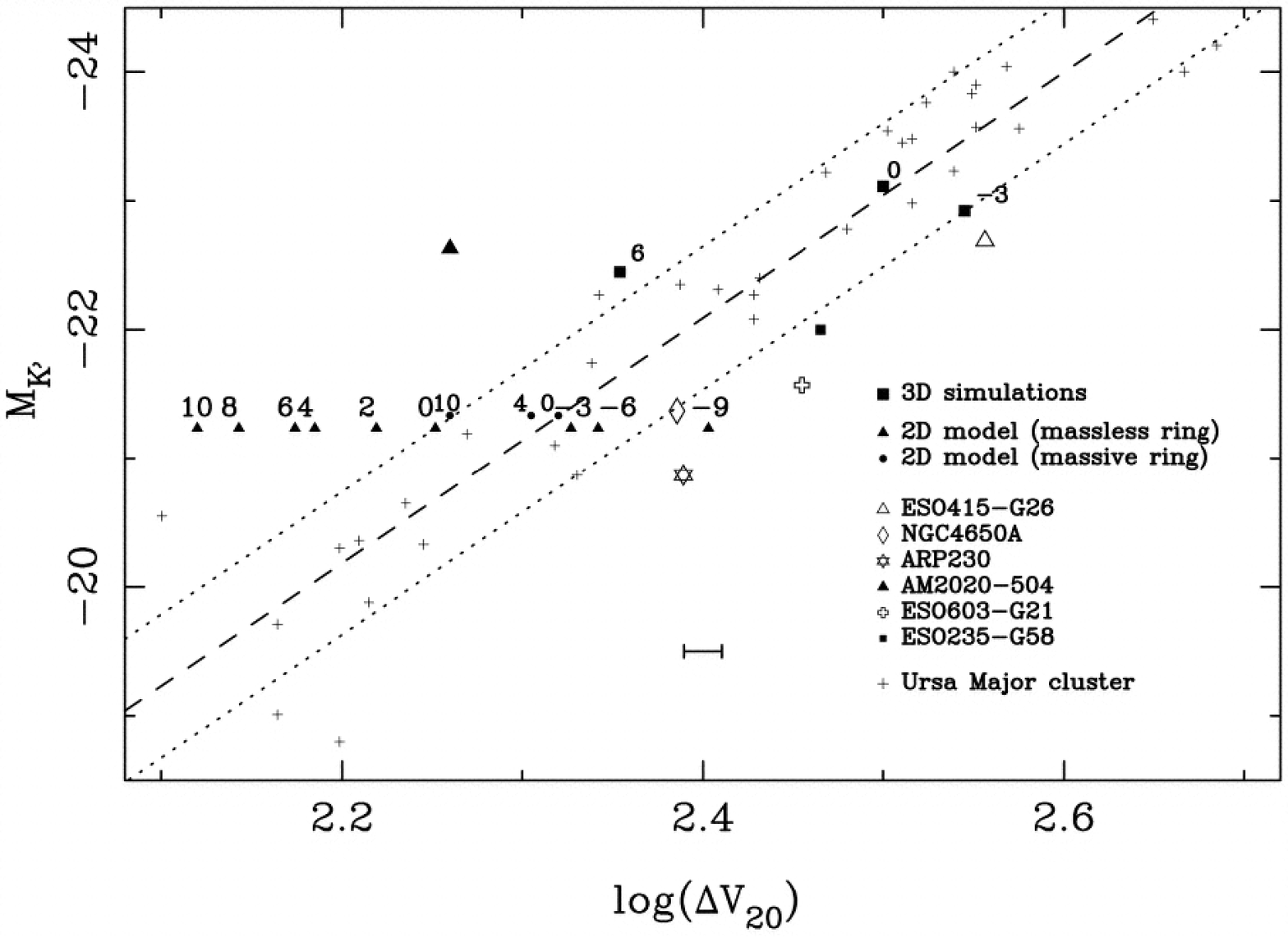}{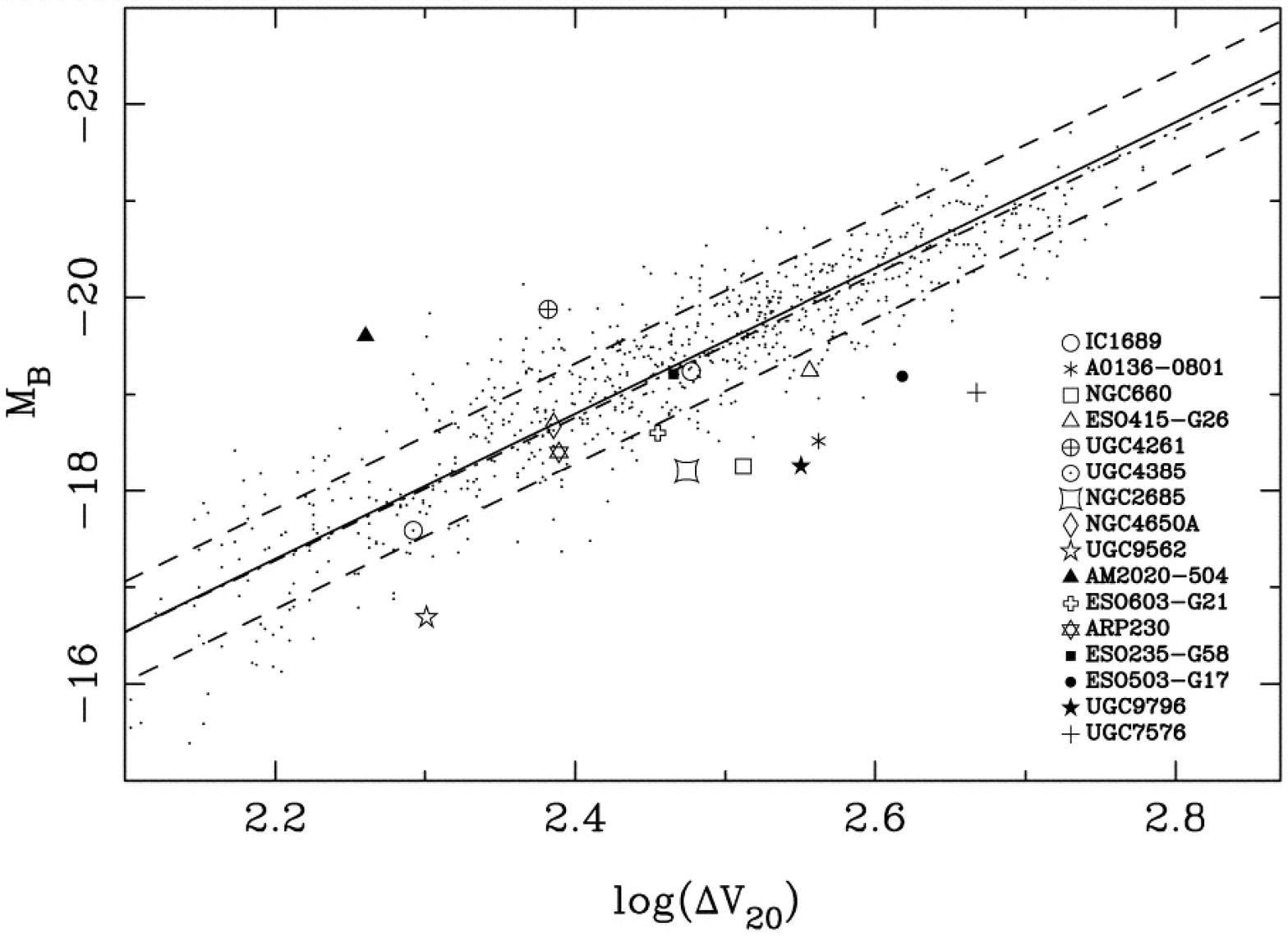}
\caption{\small Left Panel $-$ Absolute magnitude in the K band vs. the
measured \mbox{H\,{\sc i}} linewidth at $20\%$ of the peak line flux
density ($\Delta V_{20}$), for PRGs of the selected sample, compared
with a sample of spiral galaxies from Verheijen (2001), and with the
results from 3-D simulations and 2-D models (for massless rings and
rings that are as massive as the host galaxy), see also Bournaud \&
Combes (2003). The long-dashed line is
a linear interpolation of the TF relation for spiral galaxies, and the
short-dashed lines show the width at 15\% of the peak of the
statistical distribution of spiral galaxies. For both models, the
flattening of the halo is indicated next to each circle (massless
ring) or triangle (very massive ring): a positive $x$ number indicates
that it is an E$x$ halo with an equatorial flattening, while $-x$
corresponds to an E$x$ halo flattened toward the polar plane. The
results from the 3-D models shown in this plot are those computed for
the accretion scenario (Reshetnikov \& Sotnikova, 1997) ; our values
for $\Delta V_{20}$ vs. M$_K$ are very similar when one considers the
merging scenario (Bekki, 1998). Right Panel $-$ Absolute magnitude in
the B band vs. the linewidth at $20\%$ of the peak line flux density
($\Delta V_{20}$), for 15 PRGs.  Data for disk galaxies (dots) are
from Giovanelli et al. (1997).  Absolute magnitudes have been
normalized to the same value of $H_0$ for PRGs and disk galaxies ($75$
km s$^{-1}$ Mpc$^{-1}$).  A linear interpolation of the TF relation is
shown for these disk galaxies (solid line), $81\%$ of which lie inside
the dashed lines that are computed at 25\% of the peak of the
statistical distribution of spiral galaxies. The long-short dashed
line is obtained for unbarred disks, seen nearly edge-on
($i\geq80°$). From Iodice et al. (2003).}
\end{figure}  

\subsection{PRGs and the TF Relation for Spiral Galaxies} 
In Fig.~1 we show the K-band TF relation for a sample of spiral
galaxies studied by Verheijen (2001).  The values of $M_{K}$ and
$\log(\Delta V_{20})$ for PRGs are also shown on this plot.  Our and
Verheijen's data sets have very similar photometric properties and
limiting magnitudes. We see that five PRGs lie near to the
high-velocity boundary of the TF relation, or show larger velocities
(for a given luminosity) than disk galaxies. Only the PRG AM2020-504
shows a lower rotation velocity for its K-band absolute magnitude.

In the B-band, the tendency of PRGs to have larger velocities with
respect to the TF is confirmed for a larger sample of PRGs compared
with data for 787 disk galaxies (Giovanelli et al. 1997). In Fig.~1,
we see that two PRGs lie at lower velocities than those predicted by
the TF for spiral galaxies, two objects lie on the TF relation, and
twelve PRGs either lie on the high velocity boundary of the TF
relation or show much larger velocities. Van Driel et al. (2003) also
find that most kinematically confirmed PRGs show larger \mbox{H\,{\sc
i}} profile widths than bright spiral galaxies, at a given luminosity.
 
\section{The PRG positions in the $\log(\Delta V) -L$ plane 
and their implications for the dark halo} 
 
Observing higher or lower velocities with respect to the linear TF of 
disk galaxies is relevant for the discussion on the dark halo shape.
Via analytical models and simple assumptions about the mass distribution, 
either luminous or dark, we can estimate where the PRGs ought to lie
in the $\log(\Delta V)-M_{K/B}$  plane (Fig.~1),
with respect to the TF relation for disk galaxies.
If there were no dark matter, and the gravitational potential 
were oblate in the same sense as the flattened host galaxy, the polar ring 
would acquire an eccentric shape. When the polar ring 
and the host galaxy are both seen edge-on, which is close to being the case
for most of our PRGs, the net effect will be that the line-of-sight (LOS)
polar ring velocities are reduced.
In the logarithmic, scale-free, 
potential case, a simple formula gives the expected velocity ratio between 
the major and the minor axis components as:
\begin{equation}
\epsilon_v = 1 - \frac{v_{minor}}{v_{major}} = 
\epsilon_\rho \simeq 2\epsilon_\Phi
\end{equation}
from Gerhard \& Vietri (1986), where $\epsilon_\rho$ is the flattening 
(1-axis ratio) of the density distribution and $\epsilon_\phi$ is the 
potential flattening. 
Therefore we would expect PRGs to have on average lower velocities with 
respect to what would be measured in the equatorial plane.

\begin{figure}[h]
\plottwo{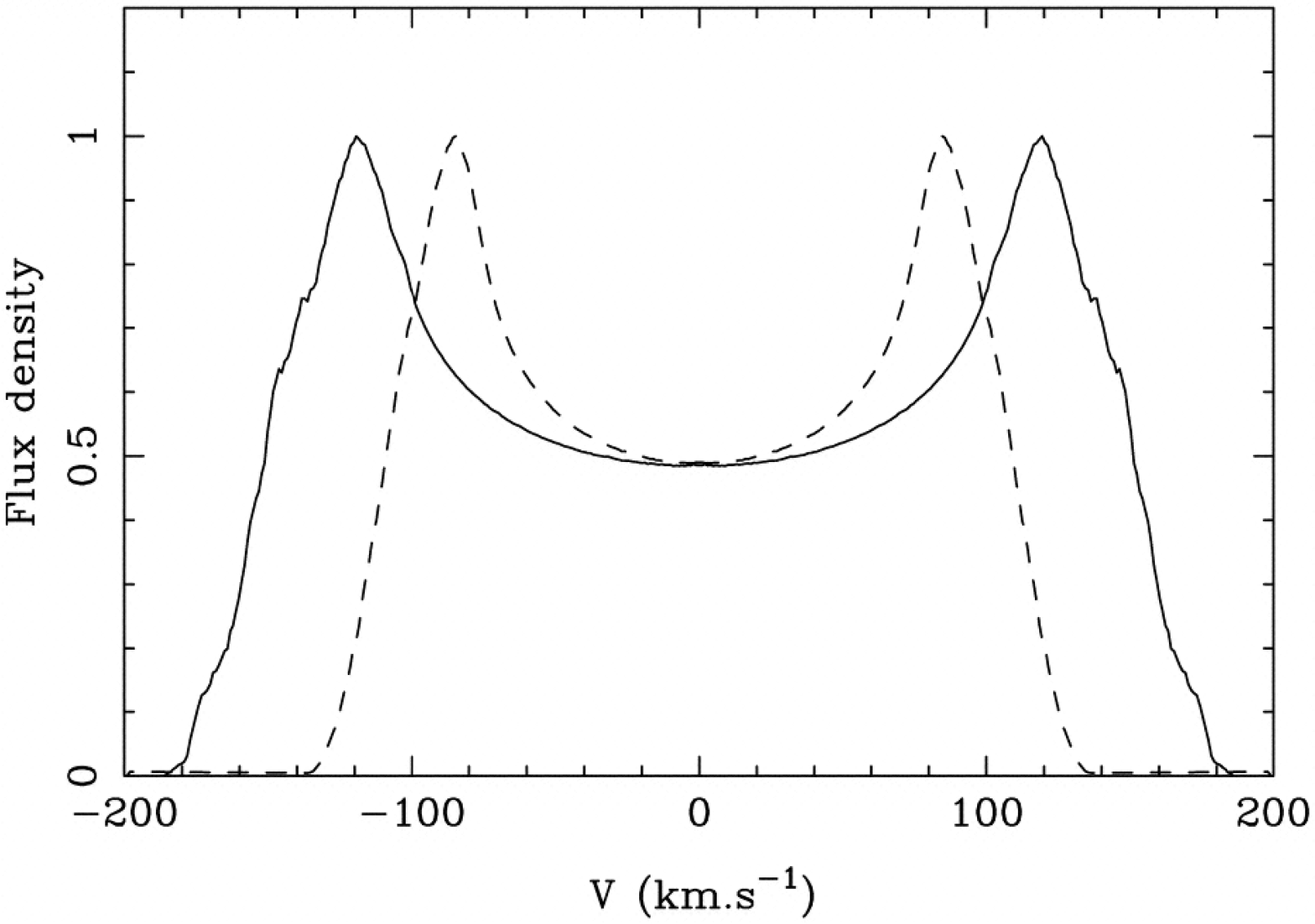}{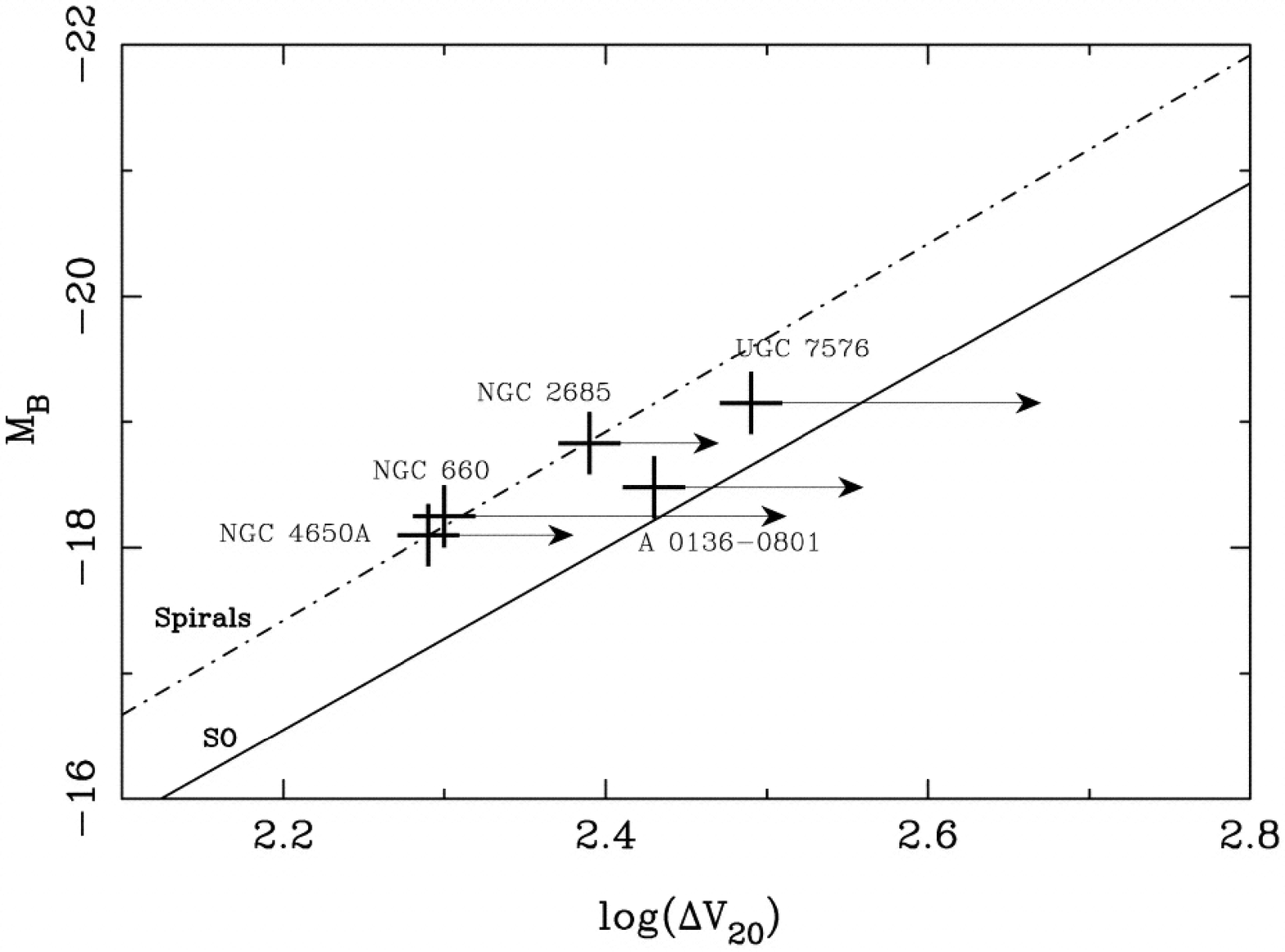}
\caption{\small Left panel $-$ Simulated \mbox{H\,{\sc i}} profiles
for circular and eccentric polar rings seen edge-on.  Solid line:
\mbox{H\,{\sc i}} profile computed for a circular ring, extending from
10 to 15 kpc. Dashed line: \mbox{H\,{\sc i}} profile computed for an
eccentric ring, with the same radial extension and radius, and
ellipticity 0.35.  The values of $\log(\Delta V_{20})$ are 2.51 for
the circular ring and 2.38 for the eccentric ring.  This difference is
in agreement with the set of values for $\Delta V_{20}$, at a given
total K or I band luminosity, obtained from PRG numerical
models. Right panel $-$ TF relation for S0 galaxies (data from Mathieu
et al. 2002) and PRGs.  The TF for spiral galaxies is plotted as a
dashed line.  Large crosses show the position of the central component
in five PRGs. From Iodice et al. (2003).}
\end{figure} 

This implies that when the polar structure is eccentric, the observed
LOS velocities in $\Delta V_{20}$ are the smallest, i.e. those from
the particles in the polar regions, see Fig.~2.  Thus the observed
$\Delta V_{20}$ depends on both the mean velocity along the ring, and
the ring eccentricity.  On the contrary, Fig.~1 shows that the
majority of PRGs have larger velocities than expected in the
$\log(\Delta V)-M_{K/B}$ plane. Therefore we need to investigate how
these velocities can be produced, and how they may depend on the
intrinsic properties of the dark galaxy halo. We computed a series of
N-body models of the formation of PRGs (Bournaud \& Combes 2003). Our
3-D and 2-D simulations of PRGs show different positions in the TF
plane depending on the shape of their dark halos. When the halo is
oblate and flattened towards the host galaxy, the observed velocity in
the polar ring are then smaller, and PRGs lie on the left-hand of the
TF relation for bright disks. When the dark halo is flattened towards
the polar ring plane, the observed velocities as larger, shifting the
PRGs to the right-hand side of the diagram as in Fig.~1 (see also
Iodice et al. 2003).

\section{Discussion}

In the $\log(\Delta V) - M_{K/B}$ planes shown in Fig.~1,
most PRGs have larger \mbox{H\,{\sc i}} rotation velocities than
standard spiral galaxies, at a given K or B-band luminosity of the
stellar component.  Our N-body simulations have suggested that a
likely explanation for this effect is a flat dark halo, whose main
plane is aligned with the host galaxy meridian plane, and prevents the
polar ring to become eccentric.  The question arises if other effects,
i.e. non-homogeneities in the TF relations for spirals, caused by bar
and/or non edge on disks, or larger M/L ratios, can produce similar
results and therefore be alternative explanations for the high
velocities observed in PRGs. 

Can the offset between the TF relation for bright spirals and PRGs be
caused by the PRGs being less luminous than spirals at a given
velocity?  No, it cannot: as shown in Fig.~2 the host galaxies in 5
PRGs fall on the TF relation for bright spirals, which indicates 1)
that the M/L ratio of this component is different from those of
standard S0s, and 2) a luminous-to-dark matter content ratio similar
to those of standard bright disks.  Gerhard et al. (2001) showed that
elliptical galaxies follow a TF relation in the $\log(\Delta V) -
M_*$, where $M_*$ is the total mass in the luminous component, which
is shallower than the relation for spiral galaxies, even when the
maximal $M/L_B$ is adopted to compute the total stellar masses.  This
led Gerhard and collaborators to infer that elliptical galaxies have
slightly lower baryonic mass than spiral galaxies of the same circular
velocities, and that their dark halos are denser than halos of spiral
galaxy with the same $L_B$.  How much more massive must the dark halo
be, to account for the velocities observed in polar rings? The
observed value of $\Delta V_{20}$ depends largely on the dark halo
shape and the ring eccentricity, while it varies only as the square
root of the total mass (and depends even less on the dark mass).
Thus, a large amount of dark matter is needed when the halo is not
polar: it ranges from factor 2 for a spherical halo, to a factor 3.5/4
for an oblate/prolate halo. Such a massive halo would cause a large
offset of the host galaxy from the TF relation of bright disks, which
is not observed.

\acknowledgments{M.A. wishes to thank the SOC for such a stimulating
Symposium, and gratefully acknowledge the IAU for financial support.}

\end{document}